\pdfoutput=1

\documentclass[preprint,12pt]{elsarticle}

\usepackage{graphicx} \usepackage{amssymb,epstopdf,amsmath,multirow}
\usepackage{setspace}
\usepackage{multirow}
\usepackage{array}
\usepackage{booktabs}

\begin{document}

\newcommand{\lanlphys}{P-23, Los Alamos National Laboratory, Los Alamos, NM 87545} 
\newcommand{\gam}{$\gamma$} \newcommand{\nnprime}{(n,n$^\prime\gamma$)} 
\newcommand{\nonubb}{0$\nu \beta \beta$}
\newcommand{\sege}{$^{68}$Ge} \newcommand{\ssge}{$^{76}$Ge} \newcommand{\ssga}{$^{76}$Ga} 
\newcommand{\sega}{$^{68}$Ga} \newcommand{\natge}{$^{\textrm {nat}}$Ge}
\newcommand{\ege}{$^{\textrm {enr}}$Ge}
\newcommand{\sco}{$^{60}$Co} \newcommand{\sfga}{$^{65}$Ga}
\newcommand{\mj}{M{\sc{ajorana}}} \newcommand{\mjdem}{M{\sc{ajorana}} D{\sc{emonstrator}}}
\newcommand{\unit}{ct t$^{-1}$ y$^{-1}$ ROI$^{-1}$}
\newcommand{\mbb}{$\langle m_{\beta\beta} \rangle$}


\title{Depth Requirements for a Tonne-scale \ssge\ Neutrinoless \hbox{Double-beta Decay} Experiment}

\address[alberta]{Centre for Particle Physics, University of Alberta, Edmonton, AB, Canada}
\address[blhill]{Department of Physics, Black Hills State University, Spearfish, SD, USA}
\address[ITEP]{Institute for Theoretical and Experimental Physics, Moscow, Russia}
\address[JINR]{Joint Institute for Nuclear Research, Dubna, Russia}
\address[lbnl]{Nuclear Science Division, Lawrence Berkeley National Laboratory, Berkeley, CA, USA}
\address[lanl]{Los Alamos National Laboratory, Los Alamos, NM, USA}
\address[queens]{Department of Physics, Queen's University, Kingston, ON, Canada}
\address[uw]{Center for Experimental Nuclear Physics and Astrophysics, and Department of Physics, University of Washington, Seattle, WA, USA}
\address[uchic]{Department of Physics, University of Chicago, Chicago, IL, USA}
\address[unc]{Department of Physics and Astronomy, University of North Carolina, Chapel Hill, NC, USA}
\address[duke]{Department of Physics, Duke University, Durham, NC, USA}
\address[ncsu]{Department of Physics, North Carolina State University, Raleigh, NC, USA}
\address[ornl]{Oak Ridge National Laboratory, Oak Ridge, TN, USA}
\address[ou]{Research Center for Nuclear Physics and Department of Physics, Osaka University, Ibaraki, Osaka, Japan}
\address[pnnl]{Pacific Northwest National Laboratory, Richland, WA, USA}
\address[sdsmt]{South Dakota School of Mines and Technology, Rapid City, SD, USA}
\address[usc]{Department of Physics and Astronomy, University of South Carolina, Columbia, SC, USA}
\address[usd]{Department of Earth Science and Physics, University of South Dakota, Vermillion, SD, USA}
\address[ut]{Department of Physics and Astronomy, University of Tennessee, Knoxville, TN, USA}
\address[tunl]{Triangle Universities Nuclear Laboratory, Durham, NC, USA}
\address[altucb]{Alternate address: Department of Nuclear Engineering, University of California, Berkeley, CA, USA}

\cortext[cor1]{Corresponding author}

\author[pnnl]{E.~Aguayo} 
\author[usc,ornl]{F.T.~Avignone~III}
\author[ncsu,tunl]{H.O.~Back} 
\author[ITEP]{A.S.~Barabash}
\author[lbnl]{M.~Bergevin} 
\author[ornl]{F.E.~Bertrand}
\author[lanl]{M.~Boswell} 
\author[JINR]{V.~Brudanin}
\author[duke,tunl]{M.~Busch}	
\author[lbnl]{Y-D.~Chan}
\author[sdsmt]{C.D.~Christofferson} 
\author[uchic]{J.I.~Collar}
\author[ncsu,tunl] {D.C.~Combs} 
\author[ornl]{R.J.~Cooper}
\author[lbnl]{J.A.~Detwiler}
\author[uw]{P.J.~Doe}
\author[ut]{Yu.~Efremenko}
\author[JINR]{V.~Egorov}
\author[ou]{H.~Ejiri}
\author[lanl]{S.R.~Elliott}
\author[duke,tunl]{J.~Esterline}
\author[pnnl]{J.E.~Fast}
\author[uchic]{N.~Fields} 
\author[unc,tunl]{P.~Finnerty}
\author[unc,tunl]{F.M.~Fraenkle} 
\author[lanl]{V.M.~Gehman}
\author[unc,tunl] {G.K.~Giovanetti} 
\author[unc,tunl]{M.P.~Green}  
\author[usd]{V.E.~Guiseppe}	
\author[JINR]{K.~Gusey}
\author[alberta]{A.L.~Hallin}
\author[ou]{R.~Hazama}
\author[unc,tunl]{R.~Henning}
\author[lanl]{A.~Hime}
\author[pnnl]{E.W.~Hoppe}
\author[sdsmt]{M.~Horton} 
\author[sdsmt] {S. Howard} 
\author[unc,tunl]{M.A.~Howe}
\author[uw]{R.A.~Johnson} 
\author[blhill]{K.J.~Keeter}
\author[pnnl]{M.E.~Keillor}
\author[usd]{C.~Keller}
\author[pnnl]{J.D.~Kephart} 
\author[lanl]{M.F.~Kidd}	
\author[uw]{A. Knecht}	
\author[JINR]{O.~Kochetov}
\author[ITEP]{S.I.~Konovalov}
\author[pnnl]{R.T.~Kouzes}
\author[pnnl]{B.D.~LaFerriere}
\author[lanl]{B.H.~LaRoque}	
\author[uw]{J. Leon}	
\author[ncsu,tunl]{L.E.~Leviner}
\author[lbnl]{J.C.~Loach}	
\author[unc,tunl]{S.~MacMullin}
\author[uw]{M.G.~Marino}
\author[lbnl]{R.D.~Martin}
\author[usd]{D.-M.~Mei}
\author[pnnl]{J.H.~Merriman}
\author[uw]{M.L.~Miller} 
\author[usc,pnnl]{L.~Mizouni}  
\author[ou]{M.~Nomachi}
\author[pnnl]{J.L.~Orrell}
\author[pnnl]{N.R.~Overman}
\author[unc,tunl]{D.G.~Phillips II}  
\author[lbnl]{A.W.P.~Poon}
\author[usd]{G. Perumpilly}  
\author[lbnl]{G.~Prior} 
\author[ornl]{D.C.~Radford}
\author[lanl]{K.~Rielage}
\author[uw]{R.G.H.~Robertson}
\author[lanl]{M.C.~Ronquest}	
\author[uw]{A.G.~Schubert}
\author[ou]{T.~Shima}
\author[JINR]{M.~Shirchenko}
\author[unc,tunl]{K.J.~Snavely} 
\author[sdsmt]{V. Sobolev}  
\author[lanl]{D.~Steele\corref{cor1}}\ead{dsteele@lanl.gov}	
\author[unc,tunl]{J.~Strain}
\author[usd]{K.~Thomas}		
\author[JINR]{V.~Timkin}
\author[duke,tunl]{W.~Tornow}
\author[ITEP]{I.~Vanyushin}
\author[ornl]{R.L.~Varner}  
\author[lbnl,altucb]{K.~Vetter} 
\author[unc,tunl]{K.~Vorren} 
\author[unc,tunl,ornl]{J.F.~Wilkerson}    
\author[uw]{B.A. Wolfe}	
\author[JINR]{E.~Yakushev}
\author[ncsu,tunl]{A.R.~Young}
\author[ornl]{C.-H.~Yu}
\author[ITEP]{V.~Yumatov}
\author[usd]{C.~Zhang}				
			

\begin{abstract}
Neutrinoless double-beta decay experiments can potentially determine the Majorana or Dirac nature 
of the neutrino, and aid in understanding the neutrino absolute mass scale and hierarchy. Future 
\ssge-based searches target a half-life sensitivity of \hbox{$>$10$^{27}$ y} to explore the inverted 
neutrino mass hierarchy. Reaching this sensitivity will require a background rate of 
$<$1 count tonne$^{-1}$ y$^{-1}$ 
in a 4-keV-wide spectral region of interest surrounding the $Q$ value of the decay.
We investigate the overburden required to reach this background goal in
a tonne-scale experiment with a compact (copper and lead) shield based on Monte Carlo estimates of 
cosmic-ray background rates. 
We find that, in light of the presently large uncertainties in these types of calculations,  
a site with an underground depth $\gtrsim$5200 meters water equivalent is required 
for a tonne-scale experiment with a compact shield similar to the planned 40-kg \mjdem. The 
required overburden is highly dependent on the chosen shielding configuration and could be 
relaxed significantly if, for example, a liquid cryogen and water shield, or an active neutron shield
were employed. 
Operation of the \mjdem\ and GERDA detectors will serve to reduce the uncertainties on cosmic-ray 
background rates and will impact the choice of shielding style and location for a future 
tonne-scale experiment. 

Note added Apr 2013: the peer review process revealed that one of the
veto rejection factors (the factor-of-four described on p.~12) 
needs to be better established. Our re-evaluation of this
parameter to date has not yielded strong support for the value stated
in the manuscript, and it will require further study to develop a solid
estimate. In the end, these further studies will supersede this aspect
of the work described in this manuscript. 
The new study may or may not lead to the same conclusion
regarding the $\gtrsim$5200 meters water equivalent requirement for future
tonne-scale $^{76}$Ge neutrinoless double beta decay experiments. 


\end{abstract}

\maketitle

\section{Introduction}

Neutrinoless double-beta (\nonubb) decay may allow the determination 
of the Majorana or Dirac nature of the neutrino,
and aid in understanding the neutrino
absolute mass scale and hierarchy 
\citep{sch82, ell02, bar04, Eji05, avi05, avi07}. 
Guiding principles
for extending the Standard Model
do not preclude each neutrino mass eigenstate from being identical to its
antiparticle, or ``Majorana'' in nature. 
The discovery of Majorana neutrinos would have profound
theoretical implications for the extension of the Standard Model while yielding
insights into the origin of mass 
and supporting the hypothesis that leptogenesis mechanisms gave rise to the observed baryon asymmetry
\cite{Fuk86, ell04}.

\begin{figure}[h]
\begin{center}
\includegraphics [width=0.9\textwidth]{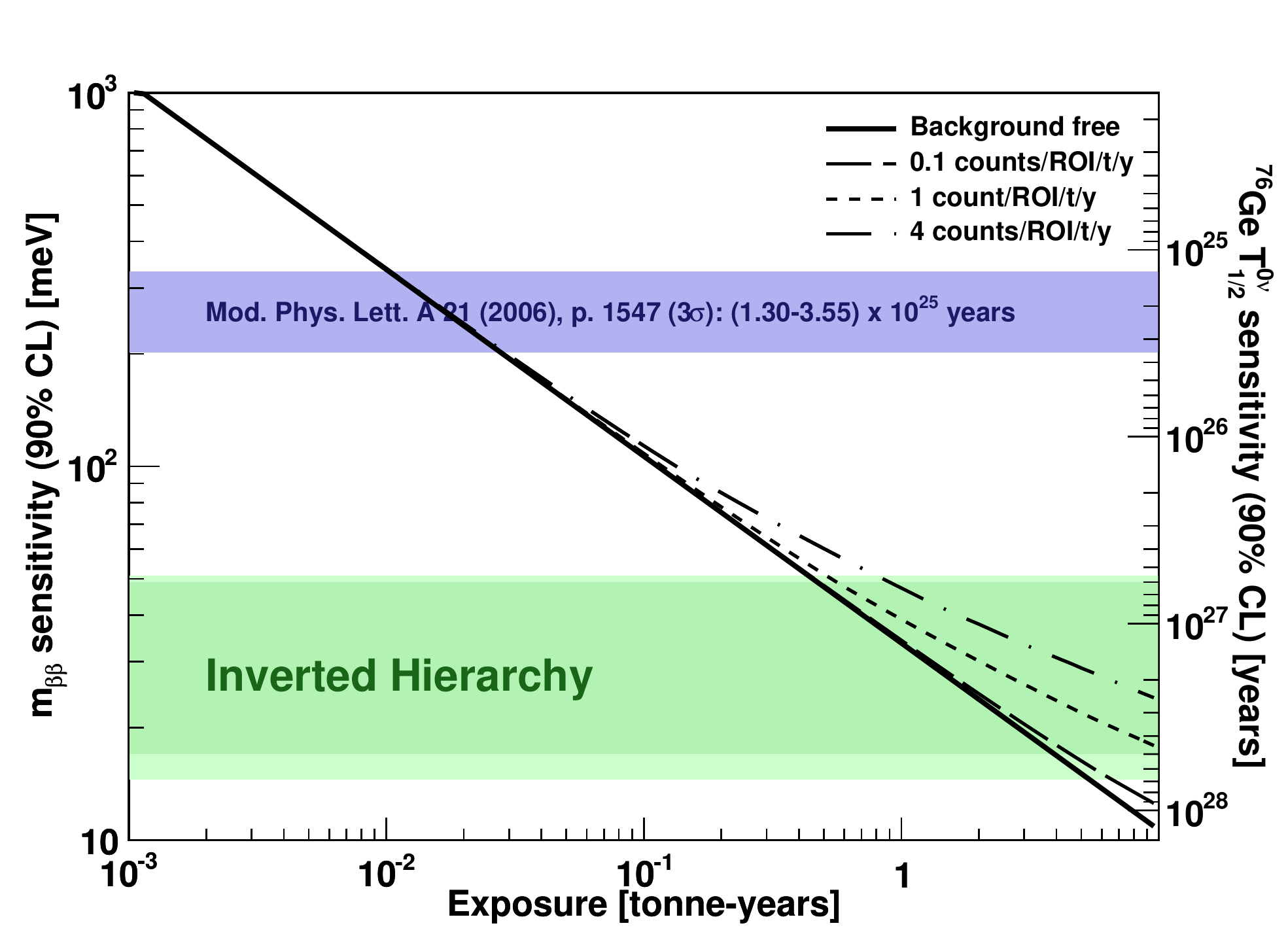}\\
\end{center}
\caption{The sensitivity of a search for \nonubb\ decay using 86\%-enriched \ssge\ as a function of background 
rate and exposure calculated according to equations 
\ref{eq:halflife} and \ref{eq:mbetabeta}. We take the signal detection efficiency, $\epsilon$, to be 68\%
to account for the effects of analysis cuts. The exposure is defined to as the product of the active
mass of the detectors and the live time of the experiment. 
The phase space factor $G^{0\nu}$ is taken from \hbox{Ref.} \cite{Suh98} and 
the conversion from half-life to \mbb\ is performed using a QRPA-calculated nuclear matrix element \citep{Sim09}.
The dark-green band indicates the range of \mbb\ in the inverted mass hierarchy for the full range 
of CP-violating phases, while the light-green bands take into account uncertainties in the measured 
neutrino oscillation parameters. The half-life corresponding to the recent claim 
of \nonubb-decay detection \cite{kla06} is shown in blue.} 
\label{fig:1TSens}
\end{figure}

Neutrino oscillation experiments \citep{ash05, ahn06, ada08} indicate that at least one neutrino mass
eigenstate has 
a mass of $\sim$45 meV or more. 
As a result,
in the inverted hierarchy mass
spectrum with $m_{3} = 0$ meV, the effective Majorana mass of the electron neutrino, 
\mbb, is between $\sim$15 and $\sim$50 meV depending on the values of the neutrino mixing parameters and Majorana phases. 
Assuming light Majorana neutrino exchange mediates the decay,
the limit on the \nonubb-decay half-life, $T^{0\nu}_{1/2}$, is 
related to the limit on the effective neutrino mass as:
\begin{equation}\label{eq:halflife}
\frac{1}{T^{0\nu}_{1/2}} = G^{0\nu} | \mathcal{M}^{0\nu} |^{2}  \langle m_{\beta\beta} \rangle ^{2}
\end{equation}
where $G^{0\nu}$ is an exactly calculable phase-space factor, and $\mathcal{M}^{0\nu}$ is a nuclear matrix element. 
The nuclear matrix element is difficult to calculate, and its value is highly model-dependent, but significant progress
in reducing the theoretical uncertainty has been made recently \citep{Rod07}. 
In the case that no signal is observed, the limit that can be placed on $\langle m_{\beta\beta} \rangle$ 
can be expressed as:
\begin{equation}\label{eq:mbetabeta}
\langle m_{\beta\beta} \rangle = \biggl[ \frac{\text{U.L.}(B)}{ln(2) N_{\text{a}} 
T \epsilon G^{0\nu}| \mathcal{M}^{0\nu} |^{2} } \biggr] ^{1/2}
\end{equation}
where \hbox{U.L.($B$)} is the upper limit on the \nonubb-decay signal given $B$ observed
background counts, $N_a$ is the number of \nonubb-decay candidate atoms, $T$ is the livetime of the experiment,
and $\epsilon$ is the signal-detection efficiency. 
As illustrated in Figure \ref{fig:1TSens}, exploring the inverted mass hierarchy by searching for \nonubb\ decay
in the isotope \ssge\ requires a sensitivity of \hbox{$T^{0\nu}_{1/2} > 10^{27}$ y}.
This is a challenging goal requiring tonne-years of exposure 
and a background rate of \hbox{$<$1 count tonne$^{-1}$ year$^{-1}$} \hbox{(ct t$^{-1}$ y$^{-1}$)} in 
the 4 keV-wide spectral region of interest (ROI) surrounding the \nonubb-decay 
$Q$ value ($Q_{\beta\beta}$) of 2039 keV, or a rate
more than a factor of 100 lower than was achieved in previous-generation experiments 
\citep{bau98,aal02}.
 
The \mj\ \cite{ell09, hen09, Aal11} and GERDA \citep{abt04, sch05} collaborations are fielding experiments to 
demonstrate the feasibility of achieving 
the background rates required for the success of a tonne-scale, \ssge-based \nonubb-decay
experiment. 
Lessons learned from both experiments will 
lead to a single, scalable design utilizing the best aspects of both technologies.  

The GERDA collaboration plans to operate (in Phase II of their experiment) a 40-kg array of enriched 
Ge (\ege) crystals immersed inside a 
cylindrical liquid argon (LAr) cryostat that is 4 m in diameter and $\sim$5.6 m tall \citep{bar09}. 
Surrounding the cryostat is a water tank 10 m in diameter and 8.9 m tall serving as a neutron moderator
and gamma-ray shield.
The water tank is also a Cherenkov detector equipped with 60 photomultiplier tubes to veto events 
coincident with through-going muons \citep{Kna09}. The collaboration is also investigating the 
possibility of instrumenting the LAr volume to serve as an inner active scintillation veto \citep{pei08}. 
The background in GERDA 
is expected to be dominated by cosmogenic and primordial radioisotope impurities
in the Ge crystals and small parts near the detectors \citep{bud09}. 

The \mjdem\ \citep{ell09, hen09, Aal11} will be an 
array of p-type, point-contact (PPC)
high-purity Ge diodes \cite{Luk89,bar07} housed in two 
vacuum cryostats built from ultra-pure electroformed Cu (EFCu).
Each cryostat will contain 35 diodes, each with mass of about 0.6 kg. 
Up to 30 kg of the detectors will be built from material enriched to at least 86\% \ssge. A graded,
passive shield will be constructed from electroformed and 
commercial high-purity Cu, high-purity Pb, polyethylene neutron moderator (some of which is borated), and
an active plastic scintillator muon veto. 
The experiment will be located in a clean 
room at the 4850-foot level (4300 meters water equivalent, \hbox{m.w.e.}) of the Sanford Underground 
Laboratory (SUL) at the Homestake mine in Lead, South Dakota. The largest sources of background in 
the \mjdem\ are expected to be the decay of cosmogenic isotopes in the Ge, primordial isotope decay 
originating in the detector mounts, cryostat and inner Cu layer of the shield, and inelastic 
scattering of muon-induced neutrons on the Ge, Cu and Pb. 

Both experiments aim to achieve a background rate of 4 \unit\ and either confirm a controversial 
claim of \nonubb-decay detection in \ssge\ \citep{kla06} 
or refute the claim and set a 
limit on the \nonubb-decay half-life. In particular, the \mjdem\ will set a limit of 
$T^{0\nu}_{1/2} > 4 \times 10^{25}$ years (90\% C.L.) after one year of running. 
In the case of a tonne-scale experiment based on the \mjdem\ concept,
additional reduction in background is achieved via more powerful analysis cuts associated with 
larger detector arrays, siting the experiment deeper underground, and if necessary, using a thicker layer 
of EFCu in the shield and fabricating the \ege\ detectors in an underground laboratory to 
reduce the amount of cosmogenic activation.\

Here we will focus on cosmic-ray backgrounds and discuss the depth requirements
for a future tonne-scale experiment. In \hbox{Sec.} \ref{sec:bgmit}, we describe the anticipated 
sources of background and strategies employed for their mitigation. We then quantify the predicted
rates of cosmic-ray backgrounds for a tonne-scale 
experiment in the context of the entire background budget, drawing upon background studies for a generic
\nonubb-decay experiment with a ``compact'' shield similar to the \mjdem,
in \hbox{Sec.} \ref{sec:depthdem}.
In \hbox{Sec.} \ref{sec:depthtonne} we derive the depth requirements for a tonne-scale experiment
and discuss the interplay between the required depth of the laboratory and the style of 
shielding.

\section{Backgrounds \& Mitigation Strategies}\label{sec:bgmit}

Achieving a low background is critical to the search for \nonubb\ decay.
In \ssge, \nonubb\ decay would produce 
a mono-energetic peak at 2039 keV. 
With a typical Ge detector resolution of 
0.2\% FWHM at 2 MeV, less than a fraction of $\sim$$2\times 10^{-15}$ of the continuous 
two-neutrino double-beta (2$\nu \beta \beta$) decay spectrum contaminates
the ROI, making this class of background completely negligible \citep{ell02}. 
Any particle with $E > Q_{\beta\beta}$ originating outside of the Ge detector volume depositing 
a portion of its energy presents a potential background. 
Even though the excellent inherent energy
resolution of Ge detectors serves to reduce the count rate in the ROI from many sources of background, 
the detectors themselves and the materials used to fabricate the detector mounts, electronic readout, 
and shielding must be of extreme radiopurity. The experiment must employ active and passive 
shielding to reduce external backgrounds originating from natural radioactivity in the cavern 
walls, and must be sited underground to reduce cosmic-ray  backgrounds.

\subsection{Depth-dependent backgrounds}\label{sec:depthdependence}

Cosmic-ray muons produce electromagnetic (EM) and hadronic particle cascades as they 
pass through rock, shielding materials, and the Ge diodes. 
Highly-penetrating neutrons with energies extending up to several 
GeV can be generated in these
cascades. In the \mjdem, these neutrons can produce backgrounds via elastic scattering on Ge,
Ge$(n,n)$, and inelastic scattering, $(n, n'\gamma)$, on Ge, Pb and Cu.
In addition, about a dozen unstable isotopes of Ge, Ga and Zn with 
half-lives greater than a few seconds and with $Q > Q_{\beta\beta}$ can be produced 
in the Ge diodes \emph{in situ} by spallation processes. 
Rarely, muons stop inside the detector and form
muonic atoms in which the muon cascades down into the 1S state and either
decays via $\mu^- \rightarrow e^- \bar{\nu_e} \nu_\mu$ or is captured by the atomic nucleus 
via the weak interaction process $\mu^- + p \rightarrow n + \nu_\mu$ \citep{mea01}. 
The reaction typically results in nuclear excitations of $\sim$100 MeV followed by the emission of 
gamma rays and a 
number of neutrons with energies typically less than 10 MeV but as high as a few tens of MeV as the 
daughter nucleus de-excites.

Neutron inelastic scattering interactions producing gamma rays coincident with muon passage through 
an anti-cosmic veto are relatively easy to eliminate.
Of greater concern are un-vetoed neutrons resulting from muon interactions in the walls and 
other components of the experimental chamber. 
While the $\mu$-induced neutron flux is typically at least three orders of magnitude
lower than the neutron flux from primordial radioactivity at underground depths greater than 
\hbox{$\sim$3000 m.w.e.} \citep{Mei06}, these neutrons are more penetrating and thus more difficult 
to shield against. Muon-induced neutron backgrounds can be suppressed by choosing a site deeper 
underground, or by minimizing the use of high-Z materials in the shield. 

As depth increases, the muon energy spectrum becomes harder and the integral flux decreases.  As a 
rule of thumb, the muon flux decreases by about an order of magnitude for every 1500 \hbox{m.w.e.} 
increase in depth \cite{Mei06}. The
integral muon flux at depths greater than $\sim$1 kilometer water equivalent (\hbox{km.w.e.}), 
$\phi_\mu$, can be parameterized as 
\begin{equation}\label{eq:muvsdepth} 
\phi_\mu(h) = a_0\, e^{-h/0.285} + a_1\, e^{-h/0.698} 
\end{equation}
where $h$ is the flat-overburden-equivalent depth in \hbox{km.w.e.}, 
$a_0 = 68\,\times  10^{-6}\, \text{cm}^{-2}\ \text{s}^{-1}$
and $a_1 = 2.1\times  10^{-6}\, \text{cm}^{-2}\ \text{s}^{-1}\,$\citep{Mei06}. Since the muon energy spectrum becomes harder with increasing 
depth, the $\mu$-induced neutron flux decreases
somewhat more slowly with increasing depth and can be parameterized as
\begin{equation}\label{eq:nvsdepth}
\phi_n (h) = P_0 \biggl ( \frac{P_1}{h} \biggr ) e^{-h/P_1}
\end{equation}
where $P_0 = (4.0 \pm 1.1) \times 10^{-7}$ cm$^{-2}$ s$^{-1}$ and $P_1 = 0.86 \pm 0.05$ \hbox{km.w.e.} \citep{Mei06}. 
Furthermore, since the cross section for the {\emph{in-situ}} production of cosmogenic isotopes
increases as a positive power of the muon energy, the rate of cosmogenic isotope production also
generally does not decrease as fast as the $\mu$ flux with depth. Several studies of $\mu$-induced 
isotopes in materials ranging from liquid scintillator to Pb have found that the production cross section,
$\sigma$, goes as
\begin{equation}\label{eq:isosigma}
\sigma(E_\mu) \propto E_\mu^\alpha
\end{equation}
where $E_\mu$ is the muon energy and $\alpha$ is a constant found to range between 0.50 and 0.93 
with a mean value $\langle \alpha \rangle = 0.73 \pm 0.10$ \citep{hag00}. Using this relation, one 
can show \citep{Mei06} that the rate of $\mu$-induced cosmogenic isotope production, $R_{iso}$, 
scales with depth as 
\begin{equation}\label{eq:isovsdepth}
\frac{R_{iso}(h=0)}{R_{iso}(h)} = \biggl ( \frac{4 \, \text{GeV}}{\langle E_\mu \rangle} \biggr )^\alpha \frac{\phi_\mu(h=0)}{\phi_\mu(h)}
\end{equation}
where $\langle E_\mu \rangle$ is the mean muon energy at depth $h$, which can be written as 
\begin{equation}
\langle E_\mu \rangle = \frac{\epsilon_\mu(1 - e^{-bh})}{\gamma_\mu - 2}.
\end{equation}
The parameters $\epsilon_\mu$, $b$ and $\gamma_\mu$ depend on 
the muon energy spectrum in the atmosphere,
muon energy loss processes in rock, and local rock density and composition. Values provided by 
Lipari \& Stanev \cite{lip91} 
(\hbox{$b = 0.383$ / km.w.e.}, $\gamma_\mu = 3.7$ and $\epsilon_\mu = 618$ GeV) and Battistoni et \hbox{al.} 
\cite{bat98} (\hbox{$b = 0.4$ / km.w.e.}, $\gamma_\mu = 3.77$ and $\epsilon_\mu = 693$ GeV)  
both agree with the measured value of the mean muon energy at the Gran Sasso National Laboratory (LNGS) \citep{amb02}.

\subsection{External \& intrinsic backgrounds}

Significant backgrounds arise from naturally-occurring primordial
radioactivity present in materials, including in the walls of the underground cavern. 
One of the isotopes in the $^{232}$Th decay chain, 
$^{208}$Tl, is of particular concern, since it decays with the emission of a 2615-keV gamma ray, well above
$Q_{\beta\beta}$ for \ssge.
An isotope in the $^{238}$U decay chain, $^{214}$Bi, also decays with the emission of 
gamma rays with energies above $Q_{\beta\beta}$. Radon and the plate-out of its 
daughters inside the shield 
and cryostat present a variety of potential $\alpha$, $\beta$ and $\gamma$ backgrounds.
A massive bulk shield composed of radiologically clean 
material and a Rn exclusion box surrounding the detector systems has been repeatedly demonstrated 
to greatly reduce primordial gamma-ray backgrounds originating externally to the shield. 

Neutrons resulting from the decay of primordial isotopes in the cavern walls, such as from the spontaneous 
fission of $^{238}$U or from natural alpha emitters via $(\alpha,n)$ reactions, are also of concern. 
These neutrons can produce gamma rays with energies above $Q_{\beta\beta}$
through inelastic scattering and capture processes in the shielding materials or in the detectors 
themselves. With a maximum energy of $\sim$10 MeV, these neutrons can be substantially suppressed by 
incorporating a hydrogenous neutron absorber into the bulk shield.  

Great care must also be taken to minimize 
the radioactive contaminants present in the components used inside the shield. 
Copper has become a material of choice for the construction of ultra-low-background detectors 
because of its lack of naturally-occurring radioisotopes, its excellent physical properties, and the 
ability to purify it to a high degree via electrodeposition \citep{Bro95,Hop07,Hop08}. 
One must ensure
that it does not have 
significant quantities of cosmogenic radioisotopes, the most troublesome of which
is $^{60}$Co, generated by $(n,\alpha)$ reactions on $^{63}$Cu. The strategy
employed by \mj\ to mitigate this background is to electroform Cu parts underground. 

Germanium detectors have the advantage that problematic radioisotopes are effectively 
removed during the detector production process. 
The cosmogenic isotope \sege\ is removed during the enrichment of the Ge material. 
Impurities of U and Th, as well as cosmogenic $^{60}$Co, are 
effectively removed during the zone refinement and crystal pulling stages. The 
relatively long-lived cosmogenic isotopes $^{60}$Co \hbox{($T_{1/2}$ = 5.3 y)} and 
$^{68}$Ge \hbox{($T_{1/2}$ = 271 d)} are produced via cosmic-ray spallation in Ge.
With a $Q$ value of 2.8 MeV, $^{60}$Co $\beta$ decay contributes to 
the background continuum. 
Although $^{68}$Ge decays by electron capture with $Q < Q_{\beta\beta}$,
its daughter isotope, $^{68}$Ga, has a half-life of \hbox{68 minutes} and decays predominantly 
by $\beta^{+}$ emission with a $Q$ value of \hbox{2.9 MeV}.
The K- and L-shell X rays from the $^{68}$Ge decay can potentially be used to tag the subsequent 
$^{68}$Ga decay and veto any events occurring 
within the time span of a few $^{68}$Ga 
half-lives. In addition, $^{68}$Ge has a half-life much shorter than the expected $\sim$10-year 
time scale of a tonne-scale experiment. 
More direct mitigation of cosmogenic backgrounds must be achieved by carefully limiting the 
aboveground exposure of the enriched material and completed detectors to cosmic rays, or 
(more ideally) by conducting as much of the detector fabrication process as practicable in an 
underground laboratory \cite{Mei11}.

\section{Simulations \& Background Studies}\label{sec:depthdem}

In this section we synthesize the results from a number of simulations and background studies to 
estimate the total background for a tonne-scale \nonubb-decay experiment with a compact shield.
The rate of cosmic-ray backgrounds as a function of depth for \ssge-based \nonubb-decay 
experiments was studied by Mei \& Hime \cite{Mei06} with an 
extensive FLUKA- and GEANT3-based custom Monte Carlo (MC) calculation.
In \hbox{Sec.} \ref{sec:muinduced} we summarize the salient results and use the scaling relations 
in \hbox{Sec.} \ref{sec:depthdependence} to derive cosmic-ray background rates for an experiment situated
at the 4850-foot level of SUL. 
Additional results attained using the same framework but performed after the 
publication of Mei \& Hime \cite{Mei06} are used to describe how modifications in the thickness of neutron shielding and muon
veto efficiency change the results. 
In \hbox{Sec.} \ref{sec:magesims} we describe our simulations of 
a 60-kg detector module with a compact shield similar to the \mjdem\ conducted within the 
MaGe \cite{bos10} simulation framework. Based on the simulation, we use a parametric model to link material 
radiopurity goals for various shield and detector components to background counts in \nonubb-decay ROI, 
taking into account analysis cut efficiencies, to demonstrate that a rate of $\sim$1 \unit\ can be 
achieved in a tonne-scale experiment.

\subsection{Cosmic-ray muon-induced backgrounds}\label{sec:muinduced}

Simulated cosmic-ray background rates, including direct contributions from untagged muons, muon capture, 
untagged $\mu$-induced fast neutron elastic and inelastic scattering, and cosmogenic radioactive isotope 
production \emph{in situ}, have been reported by Mei \& Hime \cite{Mei06}.
Their work employed FLUKA to simulate the interactions of cosmic ray muons in the rock surrounding the 
experimental apparatus. The muon energies and angular distributions were sampled from empirical 
parameterizations. The generated spallation neutrons were propagated to the cavern walls, at which 
point they were booked in histograms to be used in event generators for a GEANT3-based simulation 
of the detector response, using Hauser-Feshback theory to calculate neutron inelastic scattering 
cross sections. The simulated experimental set-up included 
a 60-kg array of enriched Ge crystals inside a compact shield with a 10-cm-thick inner Cu 
layer followed by 40 cm of Pb, 10 cm of polyethylene, and an outer active muon veto with 90\% efficiency.
To enable an easy comparison to other experiments, the simulated set-up was situated at the depth of
LNGS (3100 \hbox{m.w.e.}, flat-overburden-equivalent). Mei \& Hime \cite{Mei06} found that
the dominant $\mu$-induced contribution to the background is from fast neutron elastic and inelastic 
scattering on Ge.
The background rates of the simulated processes are summarized in Table \ref{tab:meisim}.

Before discussing how the results scale with changes in the shield configuration, 
it is necessary at the outset to mention that 
MC estimates of $\mu$-induced background rates at great depths underground are fraught with significant
uncertainty. An uncertainty of 10-15\% on the muon flux at depth is attributed to the 
uncertainties in our knowledge of primary muon flux and energy spectrum in the atmosphere 
\citep{Mei06,Kud03}. Uncertainty in the effective depth of the underground laboratory 
results in 30\% uncertainty in the muon flux normalization for the case of the 4850-foot level of 
SUL (4300 $\pm$ 200 \hbox{m.w.e.} \cite{Mei06}).
Kudryavtsev et \hbox{al.} \cite{Kud03} compared differences in muon propagation
for two different types of rock composition (``standard rock'' and Modane rock) and found 
a 15\% change in the muon flux normalization at a depth of 5000 \hbox{m.w.e.}, but no significant 
change in the mean muon energy. Adding these uncertainties 
in quadrature results in a 37\% uncertainty on the muon flux normalization at 4300 \hbox{m.w.e.} depth.

Uncertainties on the $\mu$-induced neutron fluxes are larger. 
The $\mu$-induced neutron yield in materials 
with $Z<40$ disagrees by 40\% between GEANT4 and FLUKA and by a factor of 2 for high-Z
materials such as Pb \cite{Ara04}.
While neutron yields in FLUKA are higher than in GEANT4, comparison to experimental data indicates
that both codes under predict neutron yields in light materials, and probably do so for Pb as well 
\cite{Ara04}. We therefore include a factor-of-two uncertainty on the 
$\mu$-induced neutron yield when calculating our background rates.
Uncertainties due to neutron tracking in the Monte Carlo are typically taken to be $\sim$20\% based 
on the level of agreement between different codes \cite{Pan07}. 
Mei \& Hime \cite{Mei06} note that 
neutron backscattering off of the walls of the experimental cavern can result in 
a factor of 2 to 3 change in the neutron flux incident on the shielding depending on the exact dimensions
of the cavern and shield. Mei \& Hime \cite{Mei06} included the effects of neutron backscattering in 
their simulations for a cavern with dimensions of 30 $\times$ 6.5 $\times$ 4.5 m$^3$. We take the 
uncertainty on the neutron flux due to the effect of backscattering to be 150\%, and note that for 
a larger cavern, the contribution of backscattering to the total muon-induced background is expected 
to be smaller. 
Adding the logarithms of the uncertainties in quadrature, we find
a total uncertainty on the muon-induced neutron flux of a factor of 3.2.

With these limitations in mind, we scale the Mei \& Hime \cite{Mei06} results to reflect the $\mu$-induced 
background rates in a more realistic representation of the planned \mjdem\ shield configuration. 
The major differences are: a thicker (30-cm) layer of polyethylene neutron moderator, a 99\%-efficient 
muon veto, and an overburden corresponding to the depth of the 4850-foot level at SUL
(4300 $\pm$ 200 \hbox{m.w.e.}). 
The fraction of unvetoed direct muon hits is decreased by a factor of 10 on account of the 
more efficient anti-cosmic veto. According to \hbox{Eqs.} \ref{eq:muvsdepth} and \ref{eq:nvsdepth} 
we also gain reduction factors of 5.9 and 5.6 for the direct muon flux and $\mu$-induced 
neutron flux, respectively, on account of the greater depth. Since the ``Others'' category in 
\hbox{Table \ref{tab:meisim}}
includes {\emph{in-situ}} production of cosmogenic isotopes in addition to some processes whose rates 
decrease more quickly with depth, we conservatively use \hbox{Eq. \ref{eq:isovsdepth}} to quantify 
the depth dependence of backgrounds in that category, resulting in a  reduction 
factor of 5.3. We have assumed that the effects of changes in the energy spectrum of neutrons penetrating 
through the thicker layer of polyethylene can be neglected in comparison to the large uncertainties 
on the overall $\mu$-induced neutron flux.
Simulations performed after the publication of Mei \& Hime \cite{Mei06} using the same framework showed that 
increasing the thickness of 
the polyethylene from \hbox{10 cm} to \hbox{30 cm} reduces the $\mu$-induced fast neutron flux 
penetrating the shield by a factor of two,
and that the more efficient anti-cosmic
shield decreases the number of untagged neutrons from muons that do not cross the detector by a factor of
1.8. Even though most $\mu$-induced neutrons are accompanied by the muon itself, the detection
efficiency is much lower for those that are not, and so we also
conservatively use this factor to describe the improvement in veto efficiency for all $\mu$-induced 
neutrons. 

Muon-induced neutrons emerging
from the cavern wall are typically accompanied by electromagnetic (EM) cascades. 
The Mei \& Hime \cite{Mei06} simulation treated the generation of $\mu$-induced neutrons in the rock 
separately from their propagation into and through the detector shielding. 
Simulations of the EDELWEISS experiment, a Ge-based dark matter experiment with a compact shield design similar 
to that envisioned here, indicate that including EM cascades improves the veto efficiency by almost 
an order of magnitude \citep{hor07}. 
Given the differences in the backgrounds relevant for dark matter searches and \nonubb\ decay, 
we use this value as an upper limit and 
conservatively choose a factor of four background reduction, with a factor-of-two uncertainty,
to correct for the missing EM cascades in the Mei \& Hime \cite{Mei06} simulations. 
The corresponding uncertainty introduced on the background rates is 60\%, much smaller than the
uncertainty due to neutron yield and tracking. By including this factor, 
any depth-dependent change in neutron veto efficiency due to the $\mu$-induced neutron energy spectrum 
becoming harder with increasing depth does not significantly change the results.
The scaling factors used to modify the Mei \& Hime \cite{Mei06} results as discussed above are
summarized in \hbox{Table \ref{tab:scaling}.} 

Mei \& Hime \cite{Mei06} state that the application of analysis cuts can potentially reduce the total $\mu$-induced 
neutron background rates by a factor of $\sim$7.4. However, this factor should decrease 
as the anti-cosmic veto efficiency increases from 90\%, as assumed in Mei \& Hime \cite{Mei06}, to 99\% as assumed 
in this work. The granularity and pulse-shape-analysis (PSA) cuts (see \hbox{Sec.} \ref{sec:magesims}) are most 
effective for rejecting high-multiplicity events, but as the veto efficiency increases, the average 
multiplicity of events that do not trigger the veto decreases. 
Based on our experience simulating other backgrounds, we expect analysis cuts to contribute at least 
a factor 2.5 reduction (cut efficiency of 40\%) in the rate of $\mu$-induced neutron backgrounds.
Assuming this corresponds to an upper limit on the cut efficiency, and assuming the factor of 7.4 
from 
Mei \& Hime \cite{Mei06} corresponds to a lower limit (14\% cut efficiency), we will assume an analysis 
cut efficiency of 25\%, with an uncertainty of 60\%. 

Including the effects of analysis cuts and logarithmically adding in quadrature the uncertainties on the cut 
efficiencies (60\%), muon flux normalization (37\%), neutron yield and tracking (a factor of 3.2), 
and the scaling factors introduced to adjust the Mei \& Hime \cite{Mei06} results (60\%), the total $\mu$-induced 
background rate for an experiment with a compact shield configuration situated at 
4300 \hbox{m.w.e.} is 0.33 \unit\ with an uncertainty of a factor of $\sim$4.

\subsection{MaGe simulations}\label{sec:magesims}

MaGe is an object-oriented simulation package based on ROOT \citep{bru97} 
and the GEANT4 toolkit \citep{ago02, all06} optimized for simulations of low-background Ge detector 
arrays \citep{bos10}. 
MaGe is developed and maintained jointly by the \mj\ and GERDA
collaborations.
Simulations performed to reproduce experimental data taken with a number of different detectors and 
radioactive sources generally reproduce spectral features -- such as relative peak 
heights, peak to Compton ratios, and Compton scattering multiplicity --
to within 5-10\% \cite{bos10}. 
In cases where GEANT4 has been shown to perform inadequately when compared to data, such as 
in neutron production from cosmic-ray muons and neutron inelastic scattering \citep{mar07},
efforts are underway within the \mj, GERDA, and GEANT collaborations to correct or compensate for the 
inadequacies.

Neutrinoless double-beta decay results in the emission of two electrons, and because these 
electrons travel very short distances ($\sim$1 mm) in Ge, their energy depositions are effectively 
{\it single-site} events. In contrast, most backgrounds arise from the interaction of gamma rays in 
the detector. Gamma rays with energies at or above $Q_{\beta\beta}$, with a mean distance of centimeters
between Compton scatters in Ge, are likely to deposit only a 
portion of their energy. If the energy deposited is within a few keV of $Q_{\beta\beta}$, the
interaction represents a background to the \nonubb-decay signal. Such interactions are typically 
{\it multi-site} energy depositions. 
Multi-site events can be identified and rejected if the multiple scatters occur in more than one 
crystal in a tightly-packed array. This \emph{granularity} technique is estimated to contribute a factor
2 to 5 suppression of multi-site events, depending on the origin of the background.
Modern signal digitization techniques also allow for the discrimination between single-site 
and multi-site event topologies in a single crystal. P-type, point-contact
detectors are particularly suitable for the application of these techniques. 
Measurements indicate that PSA techniques applied to data from PPC detectors 
can reject $>$90\% of multi-site background events while retaining $>$90\% of single-site 
events \citep{coo10, Gon10}. 

We have employed the MaGe simulation package to determine the background rates 
from each major background source discussed in \hbox{Sec.} \ref{sec:bgmit} for generic \ssge-based 
\nonubb-decay experiments. The simulated geometries included 3-, 9-, 21- and 57-detector arrays of 
1.05-kg, semi-coaxial Ge crystals, and a 35-crystal array of 560-g PPC detectors. For each geometry, the Cu 
mounts, front-end read-out electronics near the detectors, and cables were simulated inside a 
Cu cryostat. The detector module was enclosed inside a compact shield of Cu and Pb 
corresponding to the baseline thicknesses for the \mjdem\ described above.  Using MC calculations 
of the detector response, 
measured or targeted levels of impurities can be linked to background rates in the 4-keV-wide ROI.
For example, in the $^{232}$Th and $^{238}$U decay chains, we focus on the two radionuclides that 
have gamma-ray lines above $Q_{\beta\beta}$: $^{208}$Tl and $^{214}$Bi.
The estimates include component- and background-specific suppression
factors, including those from the granularity and PSA cuts, and for \sega\ decays, a time-correlation cut 
corresponding to five \sege\ half lives.
The granularity cut is applied to any event in which more than one detector is triggered and 
records \hbox{$>$5 keV} of energy deposited. Since our PSA software was not
fully implemented at the time when the simulations were run, the PSA background suppression 
capabilities were estimated by applying a heuristic based on the charge-drift properties and signal 
characteristics of PPC detectors, including the impact of noise and readout bandwidth. 
PSA simulations were subsequently performed and the results were found to be consistent with the 
heuristic-based analysis.
The results for the 57-crystal array (the size envisioned for a tonne-scale experiment) are 
summarized in Table \ref{tab:bgsummary}. 

Based on the absence of clear $\alpha$ peaks in the Heidelberg-Moscow data \citep{bau98}, we 
place upper limits on the $^{238}$U and $^{232}$Th contamination in the Ge 
crystals of $<$$1 \times 10^{-15}$ g/g and $<$$4 \times 10^{-15}$ g/g, respectively, corresponding to
a background rate of $<$0.30 \unit\ from all decay processes in the two decay chains, assuming secular 
equilibrium. 
We estimate a background rate of $<$0.04 \unit\ from the decay of cosmogenic \sege\ 
and \sco\ in the Ge crystals assuming the \ege\ material is limited to a maximum of 100 d 
of sea-level-equivalent cosmic-ray 
exposure. Our estimate includes the effects of PSA and granularity cuts, and the veto of any events
occuring within five \sega\ half-lives after the observation of a K-shell X ray from the decay of \sege.
The largest contribution to the total background (0.46 \unit) is expected to originate from the 
EFCu portion of the shield and the cryostats.
Additional sources of background present inside the shield include the 
detector mounts, front-end electronics, cables, and other small parts, contributing
$<$0.14 \unit. 

By simulating different background sources under a variety of detector shielding configurations, we 
were able to
parametrically extrapolate background estimates 
for different shielding designs. By doubling the thickness of EFCu shielding 
relative to that proposed for the \mjdem, we estimate that backgrounds in a tonne-scale experiment 
from an outer, commercial Cu
shielding layer would be $\sim$0.02 \unit. In this configuration, backgrounds from natural 
radioactivity in the Pb would be negligible ($<$0.02 \unit).
The effectiveness of the shield and neutron moderator limits the background from external
gamma rays and neutrons from the cavern walls to $\sim$0.05 \unit.

The total depth-independent background rate of $\lesssim$1.1 \unit\ carries an uncertainty 
of 30\%. We note that the sum of depth-independent backgrounds leaves no room for 
depth-dependent backgrounds if the background goal of $<$1 \unit\ is to be achieved.

\section{Depth Requirements}\label{sec:depthtonne}

\begin{figure*}
\begin{center}
\includegraphics [width=0.99\textwidth]{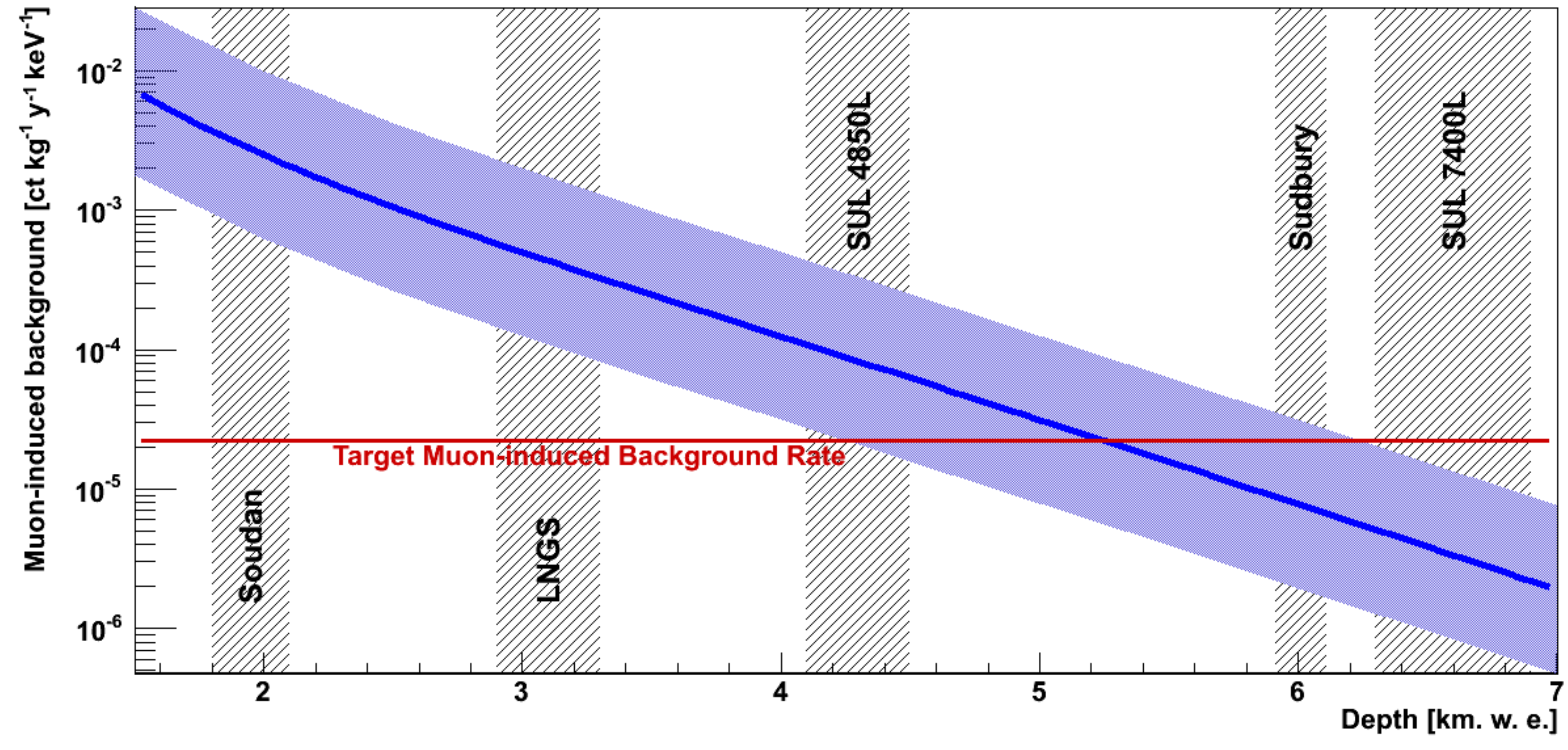} \ \ \
\end{center}
\caption{The $\mu$-induced background rate for a tonne-scale 
		\nonubb-decay experiment with a compact shield 
		 is plotted according to \hbox{Eq.} \ref{eq:muvsdepth}, normalized to a rate
         of 0.33 \unit\ at 4300 \hbox{m.w.e.} (see Table \ref{tab:meisim} and \hbox{Sec.} 
         \ref{sec:muinduced}). The blue band represents 
         a factor of 4 uncertainty in the background rate. The diagonally-shaded bands
         show the effective depths (and associated uncertainties according to \hbox{Ref.} \cite{Mei06}) 
         for several underground laboratories. The target $\mu$-induced background rate corresponds to 
         0.1 \unit. }
\label{fig:mubgvsdepth}
\end{figure*}

A tonne-scale, \ssge-based \nonubb-decay experiment must strive to make 
$\mu$-induced backgrounds negligible compared to the $\sim$1 \unit\
expected from depth-independent backgrounds. 
At a depth of 4300 \hbox{m.w.e.}, the rate of $\mu$-induced neutron backgrounds for a compact shield
configuration similar to the \mjdem\ is expected 
to be $\sim$0.3 \unit\ after analysis cuts, but with an uncertainty of a factor of 4, dominated
by uncertainties associated with the $\mu$-induced neutron yield in high-Z materials and the effects
of neutron backscattering.
A target $\mu$-induced background rate of 0.1 \unit\ implies a depth requirement of $\gtrsim$5200
\hbox{m.w.e.} assuming the expectation value of the estimated background rate. However, given the large
uncertainty, an even deeper location would be preferred to mitigate the associated risk. 
Figure \ref{fig:mubgvsdepth} illustrates this requirement in the context of several existing and 
possible future underground laboratories.

Table \ref{tab:meisim} indicates that the 
depth-dependent background for an experiment with a compact, high-Z shield configuration is dominated 
by $\mu$-induced neutron inelastic scattering on Cu and Pb components of the shield.
Eliminating high-Z materials close to the Ge detectors reduces this background contribution and can
significantly relax the depth requirement. This is the approach taken by GERDA. 
The rates of $\mu$-induced backgrounds in GERDA at the
depth of LNGS (3100 \hbox{m.w.e.}) have been calculated using a MC simulation by 
Pandola et \hbox{al.} \cite{Pan07}. The production and subsequent decay of cosmogenic 
isotopes in the Ge crystals and the LAr volume was found to be the limiting 
depth-dependent background with a rate of 
$\sim$0.8 \unit, including a conservative accounting for the effects of analysis cuts. 
The decay of $^{77}$Ge and $^{77m}$Ge produced by neutron 
capture of thermalized $\mu$-induced neutrons produced in the shield volume was found to be the dominant  
contribution.
The authors assign statistical and systematic uncertainties of $\sim$20\% and 45\%, respectively, 
to the estimated rate. The lower systematic error compared to our estimate for a compact shield 
is possible because for materials with $Z<40$, the predicted neutron yields are in closer agreement 
between different MC codes than those for Pb, and backscattering neutrons in the cavern are not expected
to contribute significantly to the background. Pandola \hbox{et al.} \cite{Pan07} also found that the $\mu$-induced 
background rates are an order of magnitude smaller in the case that liquid nitrogen is used 
in the shield in place of LAr \cite{Pan07} because the 
$\mu$-induced particle showers do not develop to the same extent as in LAr. 

The $\mu$-induced background rate in GERDA would be reduced by a factor $\sim$5 to $\sim$0.15 \unit\ 
at the depth of SUL according to \hbox{Eq.} \ref{eq:isovsdepth}. 
When comparing this rate to our estimate of 0.33 \unit\ for a compact shield, we must keep in mind 
that the uncertainty is significantly smaller, and that the estimated rate for GERDA is an upper 
limit since the assumed analysis cut efficiency for the $^{77}$Ge background is probably 
conservative \cite{Pan07}. 

The depth requirement for an experiment with a compact shield configuration is driven 
in large part by the present uncertainty associated with MC predictions of neutron yields in 
high-Z materials. The depth requirement can be relaxed significantly for a low-Z shield 
configuration since the associated $\mu$-induced background rates for low-Z configurations are 
lower and have smaller uncertainties. In particular, for a GERDA-style experiment using 
LN$_2$ for shielding, $\mu$-induced backgrounds would be negligible at the depth of the SUL 
4850-foot level. It should be noted, however, that such a shield configuration would be 
more technically complex, would require much more space, and could present significant oxygen 
deficiency and other safety hazards.

\section{Conclusions}\label{sec:conclusion}

The \mj\ and GERDA collaborations are developing experiments based on different 
shielding techniques to demonstrate that a tonne-scale, \ssge-based \nonubb-decay 
experiment can achieve the extremely low background rates necessary to accomplish the science goals. 
Achieving the desired scientific impact 
depends crucially upon reducing the background rate to $\lesssim$1 \unit.  
Even with the utmost attention to cleanliness and the radiopurity 
of construction materials, a tonne-scale experiment will likely have to contend with a 
background rate of $\sim$1 \unit\ from impurities in the detectors, their mounts and cables, 
and the shield. 
Therefore, a tonne-scale experiment should strive to make negligible all $\mu$-induced backgrounds. 

The planned depth of 4300 \hbox{m.w.e.} for the \mjdem\ is ideal in that $\mu$-induced 
backgrounds are expected to be measureable but not dominant, thus enabling 
a proper assessment of backgrounds from various different sources. 
We have shown that $\mu$-induced backgrounds are expected to 
be of significant concern for a tonne-scale experiment with a shield similar to 
that of the \mjdem\ if it were operated at the same depth.

A tonne-scale \nonubb-decay experiment with a compact shield will likely 
need to be located at a depth of $\gtrsim$5200 m.w.e. Based on our current knowledge, an 
even deeper location is preferred to mitigate the risks arising from large uncertainties in the 
estimated $\mu$-induced background rates stemming in large part from poor knowledge of $\mu$-induced 
neutron yields in high-Z materials. We note that the largest single uncertainty in the background rate
corresponds to neutron backscattering whose effects are highly dependent on the cavern and experimental 
geometry. For a cavern larger than that simulated in Mei \& Hime \cite{Mei06}, this uncertainty should
contribute to a downward change in the estimated muon-induced background rate. The depth 
requirement could also be significantly relaxed if a low-Z shielding configuration or an active neutron
veto system were employed.

The eventual choice of the location for a tonne-scale experiment 
will involve a careful analysis of the backgrounds achieved 
in the \mjdem\ and GERDA together with an evaluation of the complex interplays between the 
cosmic-ray shielding afforded by depth, technical feasibility, and background suppression capability 
of alternative shield designs.

\section*{Acknowledgments}
We acknowledge support from the Office of Nuclear Physics in the DOE Office of Science under grant 
numbers DE-AC02-05CH11231, DE-FG02-97ER41041, DE-FG02-97ER41033, DE-FG02-97ER4104, 
DE-FG02-97ER41042, DE-SCOO05054, DE-FG02-10ER41715, and DE-FG02-97ER41020. We acknowledge support 
from the Particle and Nuclear Astrophysics Program of the National Science Foundation through grant 
numbers PHY-0919270, PHY-1003940, 0855314, and 1003399. We gratefully acknowledge support from the 
Russian Federal Agency for Atomic Energy. We gratefully acknowledge the support of the U.S. Department 
of Energy through the LANL/LDRD Program. \hbox{N. Fields} is supported by the DOE/NNSA SSGF program.


\bibliographystyle{elsarticle-num}
\bibliography{majorana}

\begin{thebibliography}{10}
\expandafter\ifx\csname url\endcsname\relax
  \def\url#1{\texttt{#1}}\fi
\expandafter\ifx\csname urlprefix\endcsname\relax\def\urlprefix{URL }\fi
\expandafter\ifx\csname href\endcsname\relax
  \def\href#1#2{#2} \def\path#1{#1}\fi

\bibitem{sch82}
J.~Schechter, J.~W.~F. Valle, {Neutrinoless double-beta decay in SU(2) x U(1)
  theories}, Phys. Rev. D 25 (1982) 2951.

\bibitem{ell02}
S.~R. Elliott, P.~Vogel, {Double beta decay}, Ann. Rev. Nucl. Part. Sci. 52
  (2002) 115--151.

\bibitem{bar04}
A.~S. Barabash, {Double-beta-decay experiments: Present status and prospects
  for the future}, Phys. Atom. Nucl. 67 (2004) 438--452.

\bibitem{Eji05}
H.~Ejiri, \href{http://jpsj.ipap.jp/link?JPSJ/74/2101/}{Double beta decays and
  neutrino masses}, J. Phys. Soc. Jpn. 74 (2005) 2101--2127.

\bibitem{avi05}
F.~T. Avignone, G.~S. King, Y.~G. Zdesenko, {Next generation double-beta decay
  experiments: {M}etrics for their evaluation}, New J. Phys. 7 (2005) 6.

\bibitem{avi07}
F.~T. Avignone, III, S.~R. Elliott, J.~Engel, {Double beta decay, majorana
  neutrinos, and neutrino mass}, Rev. Mod. Phys. 80 (2008) 481--516.

\bibitem{Fuk86}
M.~{Fukugita}, T.~{Yanagida}, {Barygenesis without grand unification}, Phys.
  Lett. B 174 (1986) 45--47.

\bibitem{ell04}
S.~R. Elliott, J.~Engel, {Double beta decay}, J. Phys. G30 (2004) R183.

\bibitem{ash05}
Y.~{Ashie} {et~al.}, {Measurement of atmospheric neutrino oscillation
  parameters by Super-Kamiokande I}, Phys. Rev. D 71 (2005) 112005.

\bibitem{ahn06}
M.~H. {Ahn} {et~al.}, {Measurement of neutrino oscillation by the K2K
  experiment}, Phys. Rev. D 74 (2006) 072003.

\bibitem{ada08}
P.~{Adamson} {et~al.}, {Measurement of neutrino oscillations with the {MINOS}
  detectors in the {NuMI} beam}, Phys. Rev. Lett. 101 (2008) 131802.

\bibitem{Rod07}
V.~A. {Rodin} {et~al.}, {Assessment of uncertainties in QRPA
  $0\nu\beta\beta$-decay decay}, Nucl. Phys. A766 (2006) 107--131, errata in
  Nucl. Phys. A793 (2007) 213-215.

\bibitem{bau98}
L.~Baudis {et~al.}, New limits on dark-matter weakly interacting particles from
  the {H}eidelberg-{M}oscow experiment, Phys. Rev. D 59 (1998) 022001.

\bibitem{aal02}
C.~E. {Aalseth} {et~al.}, {{IGEX} $^{76}$Ge neutrinoless double-beta decay
  experiment: Prospects for next generation experiments}, Phys. Rev. D 65
  (2002) 092007.

\bibitem{ell09}
S.~R. Elliott {et~al.},
  \href{http://stacks.iop.org/1742-6596/173/i=1/a=012007}{The {MAJORANA}
  project}, J. Phys. Conf. Ser. 173 (2009) 012007.

\bibitem{hen09}
R.~Henning {et~al.}, The {MAJORANA DEMONSTRATOR}: An {R}\&{D} project towards a
  tonne-scale germanium neutrinoless double-beta decay search, AIP Conf. Proc.
  1182 (2009) 88--91, paper submitted to the 10th Conference on the
  Intersections of Particle and Nuclear Physics (CIPANP 2009).

\bibitem{Aal11}
C.~E. {Aalseth} {et~al.}, The {\sc majorana} experiment, Nucl. Phys. B 217
  (2011) 44, presentation at the NOW 2010 meeting.

\bibitem{abt04}
I.~Abt {et~al.}, {A new Ge-76 double beta decay experiment at LNGS},
  arXiv:hep-ex/0404039v1.

\bibitem{sch05}
S.~Sch{\"{o}}nert {et~al.},
  \href{http://www.sciencedirect.com/science/article/B6TVD-4G3CBPY-1V/2/5f8961%
e9fc5b3c476e9243ddb3845f7c}{The {GER}manium {D}etector {A}rray ({GERDA}) for
  the search of neutrinoless $\beta\beta$ decays of $^{76}${Ge} at {LNGS}},
  Nuc. Phys. B - Proc. Sup. 145 (2005) 242--245.

\bibitem{bar09}
I.~Barabanov {et~al.},
  \href{http://www.sciencedirect.com/science/article/B6TJM-4W3HX20-3/2/ab93209%
31a4e78279b83828d60af1f65}{Shielding of the {GERDA} experiment against external
  gamma background}, Nucl. Instrum. Meth. A 606 (2009) 790--794.

\bibitem{Kna09}
M.~Knapp {et~al.},
  \href{http://www.sciencedirect.com/science/article/B6TJM-4WCTWJY-4/2/edb4029%
e4c57f04f970f17885111ec61}{The {GERDA} muon veto {C}herenkov detector}, Nucl.
  Instrum. Meth. A 610 (2009) 280 -- 282.

\bibitem{pei08}
P.~{Peiffer} {et~al.}, {Pulse shape analysis of scintillation signals from pure
  and xenon-doped liquid argon for radioactive background identification},
  Journal of Instrumentation 3 (2008) 8007.

\bibitem{bud09}
D.~Budj\'{a}s {et~al.},
  \href{http://www.sciencedirect.com/science/article/B6TJ0-4VDY7XK-3/2/6899cb3%
c6b2ffd4f5ef4b96d3da0aa5c}{Gamma-ray spectrometry of ultra low levels of
  radioactivity within the material screening program for the {GERDA}
  experiment}, Appl. Radiat. Isot. 67 (2009) 755--758.

\bibitem{Luk89}
P.~Luke {et~al.}, Low capacitance large volume shaped-field germanium detector,
  IEEE Trans. Nucl. Sci. 36 (1989) 926 --930.

\bibitem{bar07}
P.~S. Barbeau, J.~I. Collar, O.~Tench, {Large-Mass Ultra-Low Noise Germanium
  Detectors: Performance and Applications in Neutrino and Astroparticle
  Physics}, JCAP 0709 (2007) 009.

\bibitem{kla06}
H.~V. Klapdor-Kleingrothaus, I.~V. Krivosheina, {The evidence for the
  observation of 0$\nu\beta\beta$ decay: The identification of 0$\nu\beta\beta$
  events from the full spectra}, Mod. Phys. Lett. A21 (2006) 1547--1566.

\bibitem{mea01}
D.~F. Measday,
  \href{http://www.sciencedirect.com/science/article/B6TVP-4465HY3-5/2/5bad296%
41e663843f2d8249b86c1541f}{The nuclear physics of muon capture}, Phys. Rep. 354
  (2001) 243--409.

\bibitem{Mei06}
D.-M. Mei, A.~Hime, Muon-induced background study for underground laboratories,
  Phys. Rev. D 73 (2006) 053004.

\bibitem{hag00}
T.~Hagner {et~al.},
  \href{http://www.sciencedirect.com/science/article/B6TJ1-40CS21S-4/2/09fc50b%
969e82e4e4909c4cb7f637a07}{Muon-induced production of radioactive isotopes in
  scintillation detectors}, Astropart. Phys. 14 (2000) 33--47.

\bibitem{lip91}
P.~Lipari, T.~Stanev, Propagation of multi-{TeV} muons, Phys. Rev. D 44 (1991)
  3543--3554.

\bibitem{bat98}
G.~Battistoni {et~al.}, {Study of photonuclear interaction of muons in rock
  with the MACRO experiment}, arXiv:hep-ex/9809006v1.

\bibitem{amb02}
M.~Ambrosio {et~al.}, {Measurement of the residual energy of muons in the Gran
  Sasso underground laboratories}, Astropart. Phys. 19 (2003) 313--328.

\bibitem{Bro95}
R.~Brodzinski {et~al.}, Low-background germanium spectrometry the bottom line
  three years later, J. Radioanal. Nucl. Chem. 193 (1995) 61--70.

\bibitem{Hop07}
E.~Hoppe {et~al.},
  \href{http://www.sciencedirect.com/science/article/B6TJM-4NG3T5H-8/2/91bc1e7%
e4334573f394f4d58ccbceafe}{Cleaning and passivation of copper surfaces to
  remove surface radioactivity and prevent oxide formation}, Nucl. Instrum.
  Meth. A 579 (2007) 486.

\bibitem{Hop08}
E.~Hoppe {et~al.}, \href{http://dx.doi.org/10.1007/s10967-008-0716-5}{Use of
  electrodeposition for sample preparation and rejection rate prediction for
  assay of electroformed ultra high purity copper for $^{232}${Th} and
  $^{238}${U} prior to inductively coupled plasma mass spectrometry
  {(ICP/MS)}}, J. Radioanal. Nucl. Chem. 277 (2008) 103--110.

\bibitem{Mei11}
D.-M. {Mei} {et~al.}, Underground high-purity germanium crystal growth for
  {DUSEL} experiment, J. Cryst. Growth, submitted (2011).

\bibitem{bos10}
M.~{Boswell} {et~al.}, {\textsc{MaGe} - a {\sc Geant4}-based Monte Carlo
  application framework for low-background germanium experiments}, IEEE Trans.
  Nucl. Sci. 58 (2011) 1212.

\bibitem{Kud03}
V.~A. Kudryavtsev, N.~J.~C. Spooner, J.~E. McMillan, {Simulations of
  muon-induced neutron flux at large depths underground}, Nucl. Instrum. Meth.
  A 505 (2003) 688--698.

\bibitem{Ara04}
H.~M. Ara\'{u}jo {et~al.}, {Muon-induced neutron production and detection with
  GEANT4 and FLUKA}, Nucl. Instrum. Meth. A 545 (2005) 398--411.

\bibitem{Pan07}
L.~Pandola {et~al.},
  \href{http://www.sciencedirect.com/science/article/B6TJM-4M9HWXX-4/2/ee2d242%
bb3fecc06897081e54132f5a5}{{M}onte {C}arlo evaluation of the muon-induced
  background in the {GERDA} double beta decay experiment}, Nucl. Instrum. Meth.
  A 570 (2007) 149--158.

\bibitem{hor07}
M.~Horn, Simulations of the muon-induced neutron background of the
  {EDELWEISS-II} experiment for {D}ark {M}atter search, Ph.D. thesis,
  Universit\"{a}t Karlsruhe (2007).

\bibitem{bru97}
R.~Brun, F.~Rademakers, {ROOT: An object oriented data analysis framework},
  Nucl. Instrum. Meth. A 389 (1997) 81--86.

\bibitem{ago02}
S.~Agostinelli {et~al.}, {{GEANT4}: A simulation toolkit}, Nucl. Instrum. Meth.
  A 506 (2003) 250--303.

\bibitem{all06}
J.~{Allison} {et~al.}, {{GEANT4} developments and applications}, IEEE Trans.
  Nucl. Sci. 53 (2006) 270--278.

\bibitem{mar07}
M.~Marino {et~al.},
  \href{http://www.sciencedirect.com/science/article/B6TJM-4PGY4K9-4/2/3428202%
6d6546477c7387b836b3457af}{Validation of spallation neutron production and
  propagation within {GEANT4}}, Nucl. Instrum. Meth. A 582 (2007) 611--620.

\bibitem{coo10}
R.~J. {Cooper} {et~al.}, A {P}ulse {S}hape {A}nalysis technique for the
  {MAJORANA} experiment, Nucl. Instrum. Meth. A 629 (2011) 303--310.

\bibitem{Gon10}
R.~Gonz\'{a}lez~de Ordu\~{n}a {et~al.}, Pulse shape analysis to reduce the
  background of {BEGe} detectors, J. Radioanal. Nucl. Chem. 286 (2010)
  477--482.

\bibitem{Suh98}
J.~Suhonen, O.~Civitarese, Weak-interaction and nuclear-structure aspects of
  nuclear double beta decay, Physics Reports 300 (1998) 123 -- 214.

\bibitem{Sim09}
F.~{{\v S}imkovic} {et~al.}, $0\nu{}\beta{}\beta{}$-decay nuclear matrix
  elements with self-consistent short-range correlations, Phys. Rev. C 79
  (2009) 055501.

\end{thebibliography}

\pagebreak


\begin{table}[h]
 \centering
 {
  \begin{tabular}{c c c} \hline \hline   
   Reaction		& \centering BG in \hbox{Ref.} \cite{Mei06} 	& Scaled BG \\
  \hline 
   \ssge$(n,n'\gamma)$		&	\centering	40			&		0.49			\\
   $^{74}$Ge$(n,n'\gamma)$	&	\centering	8.0			&		0.10			\\
   Cu$(n,n'\gamma)$			&	\centering	7.6			&		0.094			\\
   $^{208}$Pb$(n,n'\gamma)$	&	\centering	14			&		0.17			\\
   Ge$(n,n)$				&	\centering	14		    &		0.17			\\
   $\mu$ hits				&	\centering	10			&		0.17			\\
   Others					&	\centering	9.6			&	    0.13	\\ \hline
   {\bf Total}					&	\centering	{\bf 100}			&       {\bf 1.3}     \\
  \hline \hline
  \end{tabular}
  }
  \caption{\label{tab:meisim} Muon-induced background rates for a generic \nonubb-decay experiment 
  with a compact shield similar to the planned \mjdem\ operating at Gran Sasso (3100 \hbox{m.w.e.}) 
  reported in Mei \& Hime \cite{Mei06} and the same background rates scaled as described in \hbox{Sec.} 
  \ref{sec:muinduced} to represent a more realistic detector configuration at the 4850-foot level 
  (4300 \hbox{m.w.e.}) of Sanford Underground Laboratory. 
   The ``Others'' category includes muon capture, neutron capture on Ge and Cu, and prompt backgrounds
  due to cosmogenic isotopes produced {\emph{in situ}}.   Rates are given in \unit\ and do not 
  include the effects of analysis cuts. }

\end{table}

\begin{table}[h]
 \centering
 {
  \begin{tabular}{l c c c c} \hline \hline
	Background	 &	\centering Depth		& 	\centering $n$ Moderator	&	\centering $\mu$/$n$ Veto  &  Total  \\
	\hline 
   *$(n,n'\gamma)$			&	\centering	5.6	& \centering   2  &	\centering 7.2    &       81  \\
   Ge$(n,n)$				&	\centering	5.6	& \centering   2  & \centering 7.2    &       81  \\		
   $\mu$ hits				&	\centering	5.9	&	 	          & \centering 10     &       59   \\ 
   Others					&   \centering  5.3 &  \centering  2  & \centering 7.2    &       76  \\
   \hline \hline
  \end{tabular}
  }
  \caption{\label{tab:scaling} Summary of the scaling factors (described in \hbox{Sec.} \ref{sec:muinduced}) 
  used to derive the background rates shown in the right-hand column of \hbox{Table \ref{tab:meisim}}. 
  The scaling factors listed for *$(n,n'\gamma)$ apply to all neutron inelastic scattering processes in 
  \hbox{Table \ref{tab:meisim}}. The factors of 7.2 reflect the combined effects of a higher veto
  efficiency (a factor of 1.8) and accounting for the missing EM cascades accompanying muon-induced 
  neutrons emerging from the cavern walls in the simulations (a factor of 4). 
  }
\end{table}

\begin{table*}[h]
 \centering
 {\footnotesize
  \begin{tabular*}{1.0\textwidth}{@{\extracolsep{\fill}} l p{15mm} p{15mm} p{18mm} c} \toprule \midrule
   \multirow{2}{*}{\bf Source}    &   \multicolumn{3}{c}{\bf Radioactive Isotope}   &  \bf{Background}     \\ 
      &   \multicolumn{3}{c}{[\unit]}      &   [\unit]   \\ \hline\hline
  
   \multirow{2}{*}   						{  \ege\ Crystals  } 
     			& \centering    \sege  
     			& \centering    \sco  
     			& \centering     $^{232}$Th / $^{238}$U 
     & \multirow{2}{*}                        				{ $<$0.34 }  \\ 
	 & \centering 				$<$0.01		
	 & \centering 				$<$0.03  		
	 & \centering 				$<$0.30 							  
	 &   \\ \hline

   \multirow{2}{*}   						{  Detector Mounts  } 
     			& \centering    $^{208}$Tl 
     			& \centering    $^{214}$Bi   
     			& \centering     
     & \multirow{2}{*}                        				{ 0.06 }  \\ 
	 & \centering 				0.02 		
	 & \centering 				0.04  		
	 & \centering 				 							  
	 &   \\ \hline

   \multirow{2}{*}   						{  Front-end Electronics  } 
     			& \centering    $^{208}$Tl 
     			& \centering    $^{214}$Bi   
     			& \centering     
     & \multirow{2}{*}                        				{ 0.03 }  \\ 
	 & \centering 				0.01 		
	 & \centering 				0.02  		
	 & \centering 				 							  
	 &   \\ \hline

   \multirow{2}{*}   						{  Cables  } 
     			& \centering    $^{208}$Tl 
     			& \centering    $^{214}$Bi   
     			& \centering    \sco 
     & \multirow{2}{*}                        				{ $<$0.05 }  \\ 
	 & \centering 				0.02 		
	 & \centering 				0.03  		
	 & \centering 				$<$0.01 							  
	 &   \\ \hline

   \multirow{2}{*}   						{  Cryostat  } 
     			& \centering    $^{208}$Tl 
     			& \centering    $^{214}$Bi   
     			& \centering    
     & \multirow{2}{*}                        				{ 0.15 }  \\ 
	 & \centering 				0.12		
	 & \centering 				0.03 		
	 & \centering 											  
	 &   \\ \hline

   \multirow{2}{*}   						{  Inner Cu Shield } 
     			& \centering    $^{208}$Tl 
     			& \centering    $^{214}$Bi   
     			& \centering    
     & \multirow{2}{*}                        				{ 0.31 }  \\ 
	 & \centering 				0.24		
	 & \centering 				0.07 		
	 & \centering 											  
	 &   \\ \hline

   \multirow{2}{*}   						{  Outer Cu Shield } 
     			& \centering    $^{208}$Tl 
     			& \centering    $^{214}$Bi   
     			& \centering    \sco
     & \multirow{2}{*}                        				{ $<$0.03 }  \\ 
	 & \centering 				0.01		
	 & \centering 				0.01 		
	 & \centering 				$<$0.01							  
	 &   \\ \hline
	 
   \multirow{2}{*}   						{  Pb Shield  } 
     			& \centering    $^{208}$Tl 
     			& \centering    $^{214}$Bi   
     			& \centering    
     & \multirow{2}{*}                        				{ $<$0.02 }  \\ 
	 & \centering 				$<$0.01		
	 & \centering 				$<$0.01		
	 & \centering 											  
	 &   \\ \hline	 

   \multirow{2}{*}   						{  Other  } 
     			& \centering    Surface $\alpha$
     			& \centering    Rock 
     			& \centering    Other
     & \multirow{2}{*}                        				{ 0.11 }  \\ 
	  & \centering 				0.05		
	  & \centering 				$\sim$ 0.05
	  & \centering 				$\sim$ 0.01			 				  
	  &   \\ \hline	\hline
	 
   \multirow{2}{*}   						{  {\bf Total Depth-independent}  } 
     			& \centering    
     			& \centering       
     			& \centering    
     & \multirow{2}{*}                        				{ \bf $\lesssim$1.1 }  \\ 
	 & \centering 						
	 & \centering 				 		
	 & \centering 											  
	 &   \\ \hline\hline
	 
   \multirow{2}{*}   						{  {\bf Total Depth-dependent}  } 
     			& \centering    
     			& \centering       
     			& \centering     
     & \multirow{2}{*}                        				{ \bf $\sim$0.02 }  \\ 
	 & \centering 						
	 & \centering 				 		
	 & \centering 											  
	 &   \\ \hline\hline	 

   \multirow{2}{*}   						{  {\bf Total Background}  } 
     			& \centering    
     			& \centering       
     			& \centering    
     & \multirow{2}{*}                        				{ \bf $\lesssim$1.1 }  \\ 
	 & \centering 						
	 & \centering 				 		
	 & \centering 											  
	 &   \\

	\midrule 
    \bottomrule
  \end{tabular*}
  }
  \caption{\label{tab:bgsummary} Background budget for 
   a tonne-scale detector based on simulations of a 60-kg, 57-crystal detector module in a 
   compact shield. The depth-dependent background assumes the experiment is located at the 
   7400-foot level ($\sim$6600 m.w.e) of the Sanford Underground Laboratory at Homestake. 
   The total depth-independent background rate carries an uncertainty of 30\%. 
   }
\end{table*}

\end{document}